**Heuristic key to Physics beyond the Standard Model:**
**Modification of the QFTs so as to make their diagrams convergent**

Marijan Ribarič and Luka Šušteršič
Jožef Stefan Institute, p.p. 3000, 1001 Ljubljana, Slovenia
E-mail: marjan.ribaric@masicom.net

**ABSTRACT**

**Motivation:** We think that the physics of microscopic, small distance phenomena that underlie quantum scattering is the keystone of Physics beyond the Standard Model; and it may suggest cure for ultraviolet divergences, the crucial fault of QFTs.

**Objective**: Modifications of the QFTs that regularize their diagrams by an adequate account of small-distance properties of the Universe; thereby providing a heuristic key to the underlying physics of QFTs.

**Method:** A Lagrangian-based framework is used to realistically modify QFTs and make their diagrams convergent by modifying the Feynman propagators as conjectured by Pauli in 1949. The starting point is Feynman's atomistic conjecture that the partial differential equations of theoretical physics are actually describing smoothed-out, macroscopic dynamics of some infinitesimal entities. We model their microscopic dynamics by a relativistic integro-differential transport equation and use it to derive their macroscopic, large-scale dynamics.

**Results:**

a) *Testable theory:* Regularization by replacing QFT partial differential equations with relativistic integro-differential equations.

b) Novel, infinitesimal-range, fundamental forces.

c) Regularization parameters are hypothetical physical constants.

d) A classical base for explaining the QFT hierarchy problems.

e) The medium of the Universe that allows for faster than light effects.

**Application:** Collider Physics: Usage of the present high energy facilities for gathering quantitative information about the underlying physics of QFTs by the experimental values of regularization parameters.

# Content

**Footnotes:** [1] Why we need Lagrangians in field theory, see Duncan [54].

[2] A covering theory implies predictions of the original theory; cf. [4, Sect.1-2].

[3] We understand the adjectives formalistic and realistic in the sense of Pauli-Villars, cf. Sect.2.1: So we consider the regularization to be realistic if it is based on an experimentally testable hypothesis. For related definitions, see [3].

[4] We use this particular, hypothetical substance solely to facilitate organization and presentation of the heuristically needed [5] mathematical equations of real substances.

[5] Adjective heuristic describes an aid for *trial and error doings*, cf. Sect.5.2.1.

[6] Prefix micro- stands for "microscopic, small-distance".

## 1.0 Introduction

To avoid ultraviolet divergences of QFT diagrams, current calculations of quantum scattering are facilitated by formalistic[3] regularizations. Ultraviolet divergences in present QFTs seem to be a sign of unknown physics of micro-phenomena[6] of the Universe that underlie the quantum scattering; we will refer to it as *the underlying physics of QFTs.* Given a particular QFT, we propose a heuristic [5] approach to obtain a realistic[3], Lagrangian-based regularizing QFT modification of it (we will refer to as an RT) consistent with the hypothetical physical micro-laws [6] of the Universe. To elucidate them, we follow Feynman's heuristics lore [67]: "*I think equation guessing might be the best method to proceed to obtain the laws for the part of physics which is presently unknown*."

**In Sect.5.3.1 we give an overview of the basic premises**:

I.  The macroscopic, quantum phenomena of the Universe appear to be adequately modeled by the Standard Model, whereas the physics of underlying micro-phenomena6 is open, cf. [43, 75, 76, and 81]. Thus we propose to search out RTs to find information about it, cf. Sect. 5.2. We think that this physics is the keystone of Physics beyond the Standard Model. We propose to model it by the relativistic Boltzmann equations, the eight dimensional space $\mathbf{R}^{1,3} \times \mathbf{R}^{1,3}$, and a medium, which is a classical relativistic gas[4] consisting of infinitesimal fermionic and bosonic entities, which carry an arbitrary 4-momentum and interact solely by colliding, cf. Sect.4.0.

II. *A rough sketch of the entire problem,* as we see it, is as Polyakov noted in 1987 [20, Sect.12]: "Elementary particles existing in nature resemble very much excitations of some complicated medium (Aether). We do not know the detailed structure of the Aether, but we have learned a lot about effective Lagrangians for its low energy excitations. It is as if we knew nothing about the molecular structure of some liquid, but did know the Navier-Stokes equation and could thus predict many exciting things. Clearly, there are lots of different possibilities at the molecular level leading to the same low energy picture."

III. Ockham's razor is our problem-solving principle. We subscribe to Weinberg's view [45]: **a)** by taking particle physics as the culmination of

"*The ancient search for those principles*

*that cannot be explained in terms of deeper principles.*",

**b)** by writing this paper to promote that

"*...the convergence of explanations down to simpler and simpler principles will eventually come to an end in a final theory.*"

**IV.** We propose no new particles, just modified mathematics in QFTs!

**V.** Mostly, we give no explicit mathematical definition or equation, but only the corresponding references.

## 1.1 Theoretical procedures

According to Dirac [1]: "One can distinguish between two main procedures for a theoretical physicist:

**a)** One of them is to work from the experimental basis,

**b)** The other procedure is to work from the mathematical basis. One examines and criticizes the existing theory. One tries to pinpoint the faults in it and then tries to remove them. The difficulty here is to remove the faults without destroying the very great successes of the existing theory."

Ultraviolet divergences require two related procedures:

**a)** A procedure for calculating perturbative S-matrix elements. Initially physicists were searching for a radical solution – a new physical theory about the phenomena of the Universe. But calculations of perturbative S-matrix elements were eventually achieved through various formalistic [3] regularization methods.

**b)** So there is the need to obtain these excellent QFT results through a realistic [3] modification of QFTs. Of particular interest are such ones, RTs that 1) take account of experimental and theoretical constraints, and 2) facilitate information about micro-properties of the Universe.

## 1.2 The purpose of this paper

We are going to point out how to make an RT by using the Boltzmann integro-differential equations instead of QFT partial differential equations so as to avoid realistically [3] the ultraviolet divergences!

As in the last forty years there hasn't been much interest in RTs due to the utilitarian approach to QFTs by renormalization, the critical comments are forgotten, the more so, as it seems that there are no specific experimental data to check out a given RT. So it is important that we clarify the issues by collating:

- Ideas, clues and tools that might be useful.

- o    The origins of the ultraviolet divergences.
- The properties required of the RTs and their Lagrangians[1].
- Potential applications relevant to the significance of a given RT.
- Comments and viewpoints of the present-day experts and also of the Theoretical Physics founders, which Glasgow [56] pointed out:"… we were trying to help the experimenters interpret their data just as they were posing questions to us about what these strange effects they saw in the laboratory were." They might be long gone, but we feel that their remarks stay relevant!  As noted by Weinberg [74]: "Well, those were great days. The 1960s and 1970s were a time when experimentalists and theorists were really interested in what each other had to say, and made great discoveries through their mutual interchange. We have not seen such great days in elementary particle physics since that time, but I expect that we will see good times return again in a few years…"

## 1.3    Modification of the mathematical formalism of QFTs

  **I.** Instead of QFT fields of the time-space position $x \in \mathbf{R}^{1,3}$, in RTs we will use transport theoretic fields of the time-space position $x \in \mathbf{R}^{1,3}$ and the four-momentum $p \in \mathbf{R}^{1,3}$; they are multicomponent scalar, spinor, and four-vector fields of $(x, p) \in \mathbf{R}^{1,3} \times \mathbf{R}^{1,3}$. Thus, the eight-dimensional space $\mathbf{R}^{1,3} \times \mathbf{R}^{1,3}$ is the key element of an RT.

 **II.** Dimensional reduction: We express a QFT field as covariant, local average over the four-momentum $p \in \mathbf{R}^{1,3}$ of a transport theoretic field.

**III.** As equations of motion for these transport theoretic fields we use the relativistic, linearized Boltzmann integro-differential transport equations, which we call Boltzmann-Equations for short.

**IV.** We provide physical motivations for these modifications in Sect.4.0. The mathematical reasons for using Boltzmann-Equations as a tool for modifying the QFT partial differential equations we give in Sect.5.3.2.

## 2.0  Question on the present regularizations

To summarize, the present regularizations of QFTs testify to our lack of knowledge about the underlying physics of QFTs. We do not know what is happening at ultra-small distances; by renormalizations we deal with this lack without diminishing it. The key critical points on the present regularizations are listed below:

I.   **Absence of a realistic theory of quantum scattering**. Formulas of perturbative QFT are calculated using the Feynman rules with a regularization to control ultraviolet divergences and obtain convergent Feynman diagrams containing loops. The calculated n-point Green's functions and a suitable limiting procedure (a renormalization scheme) then lead to perturbative S-matrix elements. These enabled extremely well modeling of the measurable physical processes (cross sections, probability amplitudes, decay widths and lifetimes of the excited states). However, no known regularization of n-point Green's functions can be regarded as an entirely realistic theory of quantum scattering because they all disregard some basic tenets of conventional physics!

II.  **Incomplete explanation of the world**. According to Weinberg [8,p.xx]: ˝... our purpose in theoretical physics is not just to describe the world as we find it, but to explain – in terms of a few fundamental principles – why the world is the way it is.˝ Though nothing less than the single greatest intellectual achievement of the 20th century, the perturbative QFTs do not serve completely to that end due to their formalistic regularizations! This suggests looking for a realistic covering theory [2] of quantum scattering that avoids realistically ultraviolet divergences, cf. [8, Sect.1.3 and Ch.9] and [9, 10, 52, and 53].

III. **Dirac's and Feynman's criticism.** Dirac persistently questioned then available, formalistic QFT regularizations. In 1963 he wrote [11]:"... in the renormalization theory we have a theory that has defied all the attempts of the mathematician to make it sound. I am inclined to suspect that the renormalization theory is something that will not survive in the future."  So he was expecting realistically regularized theory of quantum scattering without renormalization to remove the formalistic regularization parameters! In 1990 Feynman [12] likewise wrote:"The shell game that we play is technically called 'renormalization'. But no matter how clever the word, it is still what I would call a dippy process! Having to resort to such hocus-pocus has prevented us from proving that the theory of QED is mathematically self-consistent. It's surprising that the theory still hasn't been proved self-consistent one way or the other by now; I suspect that renormalization is not mathematically legitimate."

**IV.** **The current consensus on formalistic QFT regularizations is that the low-energy behavior of Feynman propagators is adequate**. Nowadays we presume that renormalization enables us to calculate by perturbative QFT such experimentally testable approximations to the expectation values and scattering amplitudes that are the effects of the low-energy physics that seem independent of the unknown high-energy processes! A comprehensive, approving overview of renormalization is given by Gurau, Rivasseau, and Sfondrini [58].They point out that: "Quantum field theory (QFT) emerged as a framework to reconcile quantum physics with special relativity, and has now gained a central role in theoretical physics. Since its origin, QFT has been plagued by the problem of divergences, which led to the formulation of the theory of renormalization. This procedure, which initially might have appeared as a computational trick, is now understood to be the heart of QFT. In fact, the so-called renormalization group approach explains why we are able to efficiently describe complicated systems, from ferromagnetism to the Standard Model, in terms of simple theories that depend only on a small number of parameters." These remarks indicate that though the crucial usefulness of renormalization makes it the heart of the present perturbative QFTs, the theory of renormalization does not make the realistic regularizing modifications of QFTs irrelevant! In acquiring knowledge, the goals of Mathematical Physics and Theoretical Physics are quite distinct.

## 2.1 Realistic regularization

In1949 Pauli conjectured that there is *a realistic regularization*, which is obtained solely by a realistic modification of the Feynman propagators [2, 14]. The *regularization parameters* of such, realistic regularization are hypothetical physical constants that may facilitate further information about the Universe by quantum scattering. By contrast, the available regularizations introduce *formalistic regularization parameters* that have no physical significance and are eventually disposed of by renormalization. So following Pauli, the realistic regularizing modifications of a given QFT that we put forward in Sect.5 (i.e. RTs) are such that:

- RT is part of a relativistic, Lagrangian-based theory about the Universe, which adheres to the basic tenets of conventional physical theories and to the conceptual framework of QFT.
- RT parameters are considered as hypothetical physical constants, which are determined experimentally; by Glashow [56] an essential property of a physical theory.
- No knowledge about physics at unobserved scales is required!

## 3.0  Lagrangian-based modification of a QFT

Though crucial in current perturbative QFTs, the renormalizations do not satisfy the conceptual needs for *realistically regularized* QFTs! So we are going to point out a Lagrangian-based framework for constructing RTs on a conventional physical basis:

•    **In Sect.3.1** we give the reasons: the faulty high-energy asymptotes of the Feynman propagators as specified by the present QFT free-field Lagrangians.

•    **In Sect.3.2** we hypothesize a classical F-medium[4]  that propagates the QFT free fields through the Universe. In Sect.4 we specify it so that the modified QFT free-field Lagrangians ought to model the macroscopic, large-scale dynamics of the Universe better than the original ones.

•    In **Sect.3.3** we give Design Specifications for a particular RT.

## 3.1  Significance of the Feynman diagrams

Ultraviolet divergences are a serious embarrassment to theoretical physics due to the significance of the Feynman diagrams, which have transformed most of theoretical physics: they are used in the modeling calculations for all types of particle interactions. In their presentations of the fundamental interactions written from the particle physics perspective, 't Hooft and Veltman [15, 16] gave convincing arguments for taking the original, non-regularized Feynman diagrams as the most succinct representation of our present knowledge about the physics of quantum scattering of the fundamental particles. Their arguments are consistent with the following convictions of Bjorken and Drell [17]: 'The Feynman graphs and rules of calculation summarize quantum field theory in a form in close contact with the experimental numbers one wants to understand. Although the statement of the theory in terms of graphs may imply perturbation theory, use of graphical methods in the many-body problem shows that this formalism is flexible enough to deal with phenomena of non-perturbative characters. Some modification of the Feynman rules of calculation may well outlive the elaborate mathematical structure of local canonical quantum field theory."

The free-field part of the QFT Lagrangian determines the Feynman propagators, whereas the rest, i.e. the QFT interaction Lagrangian determines the vertices. As it seems that *the vertices of the original, non-regularized Feynman diagrams adequately describe interactions in quantum scattering,* it is taken that ultraviolet divergences are due to the inadequate high-energy asymptote of the Feynman propagators. Accordingly, the Pauli-Villars regularization [2], based on work by Feynman, Stueckelberg, and Rivier, modifies directly, formalistically the Feynman propagators. In this paper we point out how one can replace such formalistic modification with a realistic one, based on a hypothetical property of the Universe. Thereby we put aside an old, open question about the ultrahigh-energy processes that should be taken into account to avoid the appearance of ultraviolet divergences. Instead, we use heuristics to obtain such regularized Feynman propagators that the properties of hypothetical, micro-properties of the Universe could imply. So we start with the following hypothesis about the physical origin of the ultraviolet divergences.

## 3.2   Physical origin of the ultraviolet divergences

Within the QFT framework the ultraviolet divergences originate from summation over an infinite number of intermediate states, cf. [63, Sect. I]. We note, however, that such field-theoretic infinities were first encountered in classical electrodynamics, and then persisted in quantum electrodynamics. Thus, it makes sense to look for their origin within the framework of classical physics.

Ultraviolet divergences of momentum space integrals are due mathematically to the too slow falling at large momentum of the Feynman propagators, which are provided by the Feynman-Stueckelberg solutions to the Euler-Lagrange equations of the QFT free-field Lagrangians**.**

*We hypothesize:*

**a)**   *These Lagrangians actually model the macroscopic dynamics of an underlying F-medium[4], which propagates the QFT free fields through the Universe.*

**b)**   *To obtain Feynman propagators that avoid realistically ultraviolet divergences, these QFT free-field Lagrangians should be modified to model adequately the F-medium macroscopic dynamics!* Therefore, we adopt the following design specifications for a particular RT.

## 3.3    Design Specifications for a particular RT

The Feynman path-integral formulation provides the most direct way from the Lagrangian to the corresponding, Lorentz-invariant Feynman diagrams and also gives an insight into the physics of quantum scattering [8, Sect.1.3 and Ch.9]. So using the preceding Hypothesis and path-integral formulation, we adopt such design specifications for an RT and relations between its components as shown by the following flowchart of the regularizing modification of a given QFT.

### Flowchart of the Regularizing Modification of a given QFT

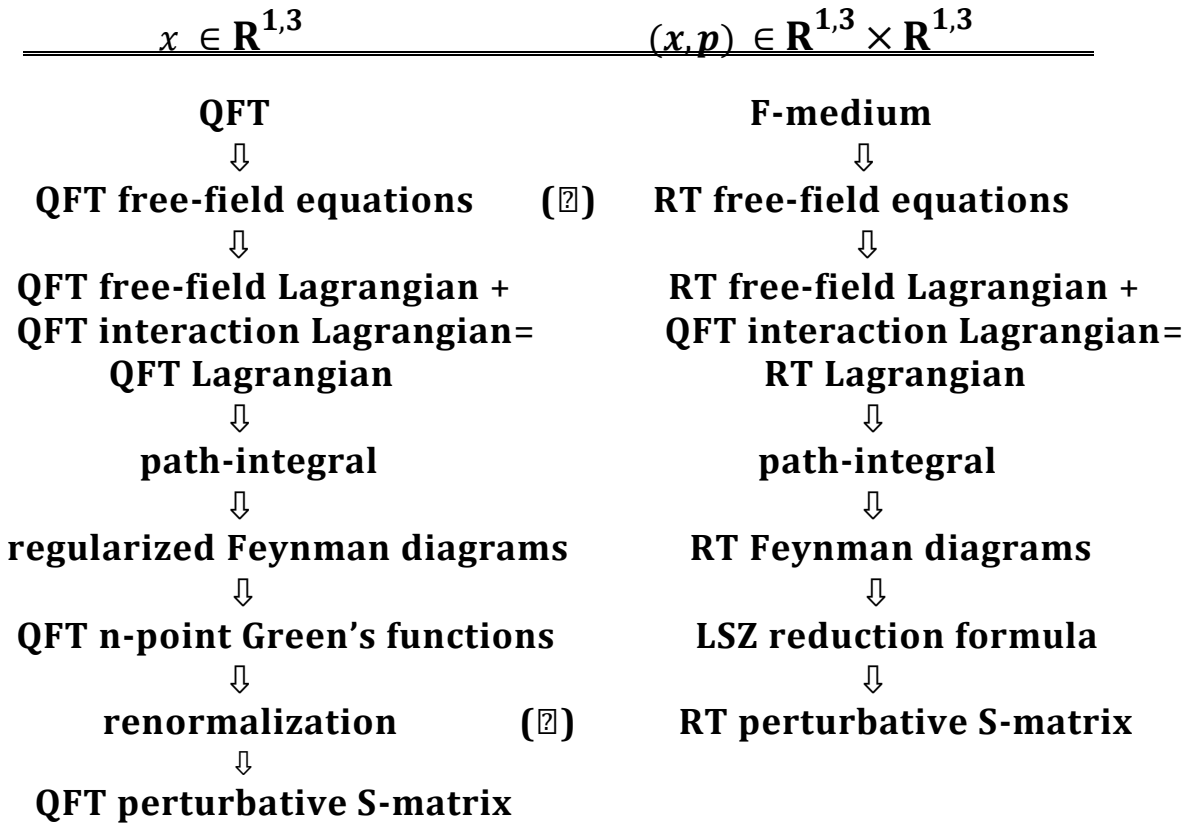

$x \in \mathbf{R}^{1,3}$                $(x,p) \in \mathbf{R}^{1,3} \times \mathbf{R}^{1,3}$

QFT                                F-medium
⇩                                  ⇩
QFT free-field equations    (⬚)    RT free-field equations
⇩                                  ⇩
QFT free-field Lagrangian +        RT free-field Lagrangian +
QFT interaction Lagrangian=        QFT interaction Lagrangian=
QFT Lagrangian                     RT Lagrangian
⇩                                  ⇩
path-integral                      path-integral
⇩                                  ⇩
regularized Feynman diagrams       RT Feynman diagrams
⇩                                  ⇩
QFT n-point Green's functions      LSZ reduction formula
⇩                                  ⇩
renormalization             (⬚)    RT perturbative S-matrix
⇩
QFT perturbative S-matrix

**RT Lagrangian** is made by retaining the QFT interaction Lagrangian, and replacing only the QFT free-field Lagrangian with one modified by the Boltzmann-Equation, a Lagrangian more precise about the macroscopic, large-scale dynamics of the F-medium, cf. Sect.4.2.2. Then we obtain by path-integral the RT Feynman diagrams. Subsequently, the Lehmann-Symanzik-Zimmermann (LSZ) reduction formula [18] provides the RT perturbative S-matrix elements.

**Euler-Lagrange equations** of the RT Lagrangian contain no formalistic parameters; their parameters are hypothetical physical constants. The original QFT Euler-Lagrange equations are their macroscopic, low-energy approximation.

**The QFT perturbative S-matrix** equals a limit of the RT perturbative S-matrix. Thus, it provides less data about the Universe. The requirements imposed on the RT Lagrangian by the symmetries of the original QFT and the unitarity of the RT perturbative S-matrix still have to be determined in detail, thereby getting also some information about micro-properties of the Universe.

## 4.0 Definition of the F-medium

Hypothesis about ultraviolet divergences in Sect.3.2 leaves open the question about the definition of the F-medium and its macroscopic dynamics whose equations model the propagation of the QFT free fields! We are going to define them in terms of the kinetic theory of gases by using Boltzmann-Equations, cf. [23, 26, 27, and 77]:

- **In Sect.4.1** we follow Feynman in 1965 and define the F-medium as a classical relativistic gas [4] consisting of infinitesimal fermionic and bosonic entities that *interact solely by colliding*.

- **In Sect.4.2**, following Bjorken and Drell in 1965, we take the view that the F-medium ultra-fast macroscopic dynamics has not been modeled adequately by using partial differential equations while making the present QFTs.

- **In Sect.4.2.2**, we therefore propose to avoid realistically ultraviolet divergences by using micro, Boltzmann-Equations in heuristic guessing adequate equations for the F-medium, ultra-fast macroscopic dynamics.

- **In Sect.5.0**, using Lagrangians of Boltzmann-Equations, we formulate the RTs, i.e. the QFTs that are modified through a realistic, Lagrangian-based regularization.

## 4.1   Underlying atomistic assumption

As stated explicitly in [57]: "A wave can be described as a *disturbance that travels through a medium from one location to another….* *To fully understand the nature of a wave, it is important to consider the medium as a collection of interacting particles…* In other words, the medium is consisting of parts that are capable of interacting with each other. The interactions of one particle of the medium with the next adjacent particle allow the disturbance to travel through the medium." Thus, we have to specify the particles [4] of the F-medium.

**Feynman's atomistic conjecture.** Commenting on the *underlying unity of the Universe*, Feynman [19,Sect.12–7] noted in 1965 that the partial differential equation of motion "... we found for neutron diffusion is only an approximation that is good when the distance over which we are looking is large compared with the mean free path. If we looked more closely, we would see individual neutrons running around." And then he wondered: "Could it be that the real world consists of little X-ons, which can be seen only at very tiny distances? And that in our measurements we are always observing on such a large scale that we can't see these little X-ons and that is why we get the differential equations? Are they [therefore] also correct only as a smoothed-out imitation of a really much more complicated microscopic world?"

The preceding statements suggest basing RTs on the following **Atomistic assumption:** *The F-medium is a relativistic gas* [4] that *consists of X-ons* [4], *which are* infinitesimal spin-0, spin-1/2, and spin-1 entities. They carry an *arbitrary four-momentum* and interact solely by colliding. So we have to specify their dynamics!

## 4.2 Dynamics of the F-medium?

*Reasons for the ultraviolet divergence by* Bjorken and Drell: Commenting on the fact that contemporary theories about the quantum scattering of fundamental particles grew out of applications of the quantization procedure to classical fields that satisfy wave equations, Bjorken and Drell [17] pointed out in 1965 the following facts about such a procedure: "The first is that we are led to a theory with differential wave propagation. The field functions are continuous functions of continuous parameters $x$ and $t$, and the changes in the fields at a point $x$ are determined by properties of the fields infinitesimally close to the point $x$. For most wave fields (for example, sound waves and the vibrations of strings and membranes) such a description is an idealization, which is valid for distances larger than the characteristic length, which measures the *granularity* of the medium. For smaller distances these theories are modified in a profound way. The electromagnetic field is a notable exception. Indeed, until the special theory of relativity obviated the necessity of a mechanistic interpretation, physicists made great efforts to discover evidence for such a mechanical description of the radiation field. After the requirement of an "ether", which propagates light waves, had been abandoned, there was considerably less difficulty in accepting this same idea when the observed wave properties of the electron suggested the introduction of a new field. Indeed there is no evidence of an ether, which underlies the electron wave. However, it is a gross and profound extrapolation of present experimental knowledge to assume that a wave description successful at "large" distances (that is, atomic lengths $\approx 10^{-8}$cm) may be extended to distances an indefinite number of orders of magnitude smaller (for example, to less than nuclear lengths $\approx 10^{-13}$cm). In the relativistic theory, we have seen that the assumption that the field description is correct in arbitrarily small space-time intervals has led—in perturbation theory—to divergent expressions for the electron self-energy and the bare charge. Renormalization theory has sidestepped these divergence difficulties, which may be indicative of the failure of the perturbation expansion. However, it is widely felt that the divergences are symptomatic of a chronic disorder in the small-distance behavior of the theory. We might then ask why local field theories, that is, theories of fields, which can be described by differential laws of wave propagation, have been so extensively used and accepted. There are several reasons, including the important one that with their aid a significant region of agreement with observations has been found. But the foremost reason is brutally simple: there exists no convincing form of a theory, which avoids differential field equations".

So following Bjorken, Drell, and Schwinger [17, 21], we may yet still wonder what kind of Lagrangians would generate adequately modified QFT free-field partial differential equations? Above Feynman's atomistic conjecture suggests that the kinetic theory of gases may provide such Lagrangians. They might enable us to modify QFTs so as to take better account of those micro-properties of the Universe that determine its macroscopic, large-scale dynamics. Thus, as suggested by Bjorken and Drell: *Though there is the renormalization theory to sidestep the ultraviolet divergences, it would be theoretically more appropriate to avoid them by* extending seamlessly the wave description to smaller distances *for more precise modeling of the Universe micro-dynamics.* On the analogy with the kinetic theory of gasses, we do this by:

I.   Using one-particle distribution functions.
II.  *Making the QFT free-field Lagrangians via Boltzmann-equations, thereby obtaining QFT free-field partial differential equations that yield realistically regularized Feynman propagators*.

## 4.2.1 Kinetic theory of gases

Gases provide a variety of media the waves travel trough. The kinetic theory of gases describes *the large-scale phenomena* by

**I. The macroscopic variables** such as *the number density, the macroscopic mean velocity, the pressure tensor, the heat-flow vector, and the energy density*, cf. [23, Sect.2.4 and Ch.3; 24, Sect.2.7 and Ch.11]. For large-scale phenomena where changes in macroscopic variables are sufficiently small over a mean collision time and over a mean free path, which is the fundamental length, one can predict the evolution of macroscopic variables accurately enough by

**II. The partial differential equations of fluid dynamics.** For a rare gas of the identical infinitesimal entities, these equations can be extended to model somewhat faster changes of the macroscopic variables by introducing additional fields of the time-space variable, which have no direct significance within the framework of the fluid dynamics, though they can be interpreted as local averages of the microscopic state, cf. the Grad method of moments [25]. (*This approach makes one wonder about the significance of certain fields in the Standard Model.*) But eventually the equations of motion cannot be improved this way anymore, and one may resort to a detailed, micro-description by

**III. The relativistic one-particle distribution function** of eight independent, continuous variables: the time-space position $x \in \mathbf{R}^{1,3}$ and the four-momentum $p \in \mathbf{R}^{1,3}$ of the constituent entities. This classical, field *determines the values of the macroscopic variables (i.e. their fields) through certain covariant, local averages over the four-momentum p,* which are counterparts to compactification in string theory.

**IV. Equation of motion** for the one-particle distribution function of a real gas is conditionally Boltzmann equation**,** which is local in the time-space variable *x,* but not in the four-momentum variable *p* [26]. Boltzmann equations take some account of the gas granularity by their linearized "molecular chaos collision term", which is a linear integral operator. Thus, they model the ultra-fast gas dynamics better than the partial differential equations of the fluid dynamics, which are such macroscopic, asymptotic approximations to Boltzmann equations that depend on the initial and boundary conditions, and neglect the micro, collision properties of gas [25, 26].

## 4.2.2  Dynamics of the F-medium by Boltzmann-Equations

Motivated by the derivation of the partial differential equations of fluid dynamics as a macroscopic, asymptotic approximation to a Boltzmann equation, and by the above Feynman's atomistic conjecture that the partial differential equations are actually describing large scale, macroscopic phenomena due to micro-motion of X-ons, we define:

**I.**  The F-medium as a gas[4] made up of infinitesimal, fermionic and bosonic entities, X-ons, which carry an arbitrary four-momentum and interact solely by colliding.

**II.**  The state of this gas is given by a four-vector one-particle distribution function, and a two-component-spinor one-particle distribution functions.

**III.**  Its Euler-Lagrange equations are Boltzmann-Equations with such collision terms that the Feynman-Stueckelberg solutions provide adequate propagators.

The macroscopic dynamics of the F-medium is completely determined by the collision term *we choose* (for examples see Sect.5.3.2, eq.(5.9); and [30, Sect.iv]). *It determines all properties of the* F-medium *we need because we actually use just the Boltzmann-equation and not the F-medium* [4] *itself!*

## 5.0  Modification of QFTs by Boltzmann-Equations

RT design specifications in Sect.3.3 and modeling of the F-medium macroscopic dynamics in Sect.4.2.2 specify a Boltzmann-Equation based framework for making RTs; cf. [27-40]**.**

Potential applications of the F-medium are mentioned in Sect.5.2. They may provide some information about appropriate collision terms of those Boltzmann-Equations we should use when modifying QFTs to get appropriate RTs.

## 5.1  Searching for an appropriate RT

*Given a QFT we may get heuristically an appropriate RT through the following five stages.*

**I.**  **Initial choice.** Ultraviolet divergences of momentum space integrals are due to the too slow falling at large momentum of the Feynman propagators. Thus the simplest RT is gotten through multiplying all of them by the same scalar valued function of momentum, "a regularizing factor" that is falling sufficiently fast at large momentum. The basic mathematical premise for selecting this regularizing factor is: **(a)** the high-energy asymptote of the Feynman propagators is wrong: and **(b)** their analytic behavior seems to be adequate elsewhere. We have yet to establish precisely the properties of a regularizing factor and of the F-medium macroscopic dynamics that are necessary for the modification of a given QFT to remain within the QFT conceptual framework and retain its properties. About the unitarity in particular, unitary regulators by 't Hooft and Veltman could provide a hint [15, 35].

**II.** **Choice of the collision term** for a physical specification of the regularizing factor, cf. [30, Sect.iv].

**III.** **Calculation of RT propagators** from the free field Lagrangians that are modified through a *Boltzmann-Equation*. Thus expressing the regularizing factor and its free parameters in the chosen collision term, cf. [30, 35, and 36].

**IV.** **Experimental evaluation** through quantum scattering**.** There is a variety of collision terms for Boltzmann-Equations that result in an RT. So the real physical significance of a particular RT must be evaluated experimentally, in determining by the well-established statistical methods: **(a)** *the values and confidence intervals of RT parameters, and* **(b)** *the increase in the accuracy of the original QFT by this RT.* This way we can also determine the significance of the new information about the Universe provided by the chosen collision term of this RT. But as Heisenberg [13] pointed out: "What we observe is not nature itself, but nature exposed to our method of questioning."

**V.** **Decision about repeating this cycle with an altered collision term,** which may be suggested by the experimental evaluation, by making of the Standard Model [74], or by constructing empirical formulas for experimental data [73].

**Note.** According to Dirac [1]:"QED is the domain of physics that we know most about, and presumably it will have to be put in order before we can hope to make any fundamental progress with other field theories, although these will continue to develop on the experimental basis."

Commenting on "The Evolution of the Physicist's Picture of Nature" and the relevant contemporary problems, Dirac [11] suggested:

"I believe separate ideas will be needed to solve these distinct problems and that they will be solved one at a time through successive stages in the future evolution of physics. At this point I find myself in disagreement with most physicists. They are inclined to think one master idea will be discovered that will solve all these problems together. I think it is asking too much to hope that anyone will be able to solve all these problems together. One should separate them one from another as much as possible and try to tackle them separately. And I believe the future development of physics will consist of solving them one at a time, and that after any one of them has been solved there will still be a great mystery about how to attack further ones." These remarks suggest that one should start by exploring all aspects of a particular kind of regularizing modifications in the case of QED. This appears the simplest, fundamental problem of theoretical physics. However, the ultimate test might be the increase of the accuracy of the Standard Model by such a modification!

### 5.2 The reasons for considering RTs

***a)*** To follow 't Hooft [6]: "History tells us that if we hit upon some obstacle, even if it looks like a pure formality or just a technical complication, it should be carefully scrutinized. Nature might be telling us something, and we should find out what it is.", cf. [88].

**b)** Ultraviolet divergences of QFTs apparently signify our inadequate treatment of the ultrahigh-energy physical processes—processes affected by the micro-properties of the Universe, cf. [8], [17, Sects. 11.2 and 11.3], and [21]. Searching out RTs, we may get information about them and so find a heuristic key to Physics beyond the Standard Model. Volovik [79] initiated by "From Topological Matter to Relativistic Quantum Vacuum" a various approach to this end.

**c)** The founders of Modern Physics conjectured: *There are regularizations of QFTs whose parameters are physical constants of the micro-properties of the Universe.*

**d)** In the following three subsections we give additional arguments for the usefulness of an RT.

### 5.2.1. Infinity problem

In quantum field theories the infinity problem was *solved* by renormalization after over forty years endeavor cf. [9, 10, 58, 74, 87]. However, according to Bjorken and Drell in 1965, the renormalization just sidestepped this problem, and as suggested by Dirac, still in early 1980s, the most important challenge in physics is to get rid of infinity yet. We propose to do this by a realistic, Lagrangian based regularization of the Feynman propagators. In 1972 Salam remarked on the skepticism over this kind of approach to ultraviolet divergences [5]:

"Field-theoretic infinities first encountered in Lorentz's computation of electron have persisted in classical electrodynamics for seventy and in quantum electrodynamics for some thirty-five years. These long years of frustration have left in the subject a curious affection for the infinities and a passionate belief that they are an inevitable part of nature; so much so that even the suggestion of a hope that they may after all be circumvented – and finite values for the renormalization constants computed is considered irrational. Compare Russell's postscript to the third volume of his Autobiography The Final Tears, 1944-1967 (George Allen and Unwin, Ltd., London 1969) p.221: 'In the modern world, if communities are unhappy, it is often because they have ignorance, habits, beliefs, and passions, which are dearer to them than happiness or even life. I find many men in our dangerous age who seems to be in love with misery and death, and who grow angry when hopes are suggested to them. They think hope is irrational and that, in sitting down to lazy despair, they are merely facing facts."

There are some, simple reasons for the above skepticism:
- As pointed out by Hossenfelder [84], it is not possible to decide whether or not such hope is justified. Each one has to decide for himself whether he is an amateur or professional, i.e. can he afford himself spending time on an arduous problem or is it *publish or perish*.
- This problem consists of instances where we are standing at the frontier of knowledge with no clear goals. So it is not obvious which information and approach are relevant.
- In such ill defined problems heuristics is a natural choice, cf. [88].

In this paper, we propose *Feynman's heuristics* to obtain a finite reformulation of the QFT theories. As Wüthrich [86] pointed out, Feynman heuristics has three distinct, major steps:

**i)**    to extend the domain of application of an existing theory,

**ii)**    to provide a model to justify the theory's equations,

**iii)**    to reveal assumptions problematic for the existing theory, and in this way find amendments to it.

Our purpose is to provide a heuristic key to Physics beyond the Standard Model; and collate utile definitions, formulae, concepts, remarks, etc.

### 5.2.2  Collider Physics

We are putting forward *the new paradigm for Collider Physics*: "Gathering information about the Physics of the Universe by the experimental values of regularization parameters of RTs."

Precision tests of the Standard Model indicate that use of the suggested heuristic key to Physics beyond the Standard Model will require extremely high precision measurements and large-scale data-intensive computing, which are customary when probing fundamental interactions, cf. [89]. According to Vos [42]:"We enter an era of exploration, where precise and sensitive measurements of known processes may be the best opportunity to reveal hints of the high-scale physics that lies beyond the Standard Model."

Outstanding results of particular RT tests as specified in Sect.5.1/v, would justify considering:

- The corresponding Boltzmann-Equation as modeling novel micro-properties of the Universe.
- RT regularization parameters as additional physical constants.
- The collision term of its *Boltzmann-Equation* as providing initial information about some novel fundamental forces.

**Benefits:**    This paradigm enables the already existing collider facilities to gather novel information about the Physics of the Universe. In view of the increasing size, complexity, and cost of the high-energy colliders [7, 62], this paradigm might be a welcome addition to the collider-based physics.

### 5.2.3  Potential applications of the F-medium

According to The Physics of the Universe report [65]:"The opportunity to gather important new knowledge in cosmology, astronomy and fundamental physics stems from recent discoveries which suggest that the basic properties of the Universe as a whole may be intimately related to the science of the very smallest known things." This might hint at a future role of the F-medium.

The F-medium is compatible with conventional theoretical physics and could provide realistic, regularizing modifications of the QFTs. However, its true significance will be determined by its future applications. There are a few potential ones we'd like to mention:

I.   **Modeling of quantum phenomena by stochastic fluctuations of the F-medium**.  In 2001 't Hooft speculated [46]: "We should not forget that quantum mechanics does not really describe what kind of dynamical phenomena are actually going on, but rather gives us probabilistic results. To me, it seems extremely plausible that any reasonable theory for the dynamics at the Planck scale would lead to processes that are so complicated to describe, that one should expect apparently stochastic fluctuations in any approximation theory describing the effects of all of this at much larger scales. It seems quite reasonable first to try a classical, deterministic theory for the Planck domain. One might speculate then that what we call quantum mechanics today, may be nothing else than an ingenious technique to handle this dynamics statistically."

II.  **Application of the F-medium to classical electrodynamics**.

III. **Modeling "dark matter" by the F-medium**, cf. [44].  Astronomical observations suggest that matter in the Universe is mainly a dark matter consisting of particles of an unknown type.

IV.  **We may regard the F-medium as the ether that is specified by the quantum scattering of fundamental particles!** It remains to adapt the F-medium properties to get a universal ether for other purposes of theoretical physics. According to the theoretical point of view (of Einstein [47], Dirac [48], Bell [49], Polyakov [20], Gorbatsevich [55], and 't Hooft [46]), there might be a non-material space filling medium enabling the observed physical processes, ether, occupying every point in space and serving as a transmission medium for the propagation of  fundamental forces, cf. [60].  L. de Broglie [59] referring to ether as subquantic medium has this to say: *"...., I have come to support wholeheartedly a hypothesis proposed by Bohm and Vigier. According to this hypothesis, the random perturbations to which the particle would be constantly subjected, and which would have the probability of presence in terms of W, arise from the interaction of the particle with a "subquantic medium" which escapes our observation and is entirely chaotic, and which is everywhere present in what we call*

*"empty space."* According to Rothwarf and Roy[ 41]:" Today the need for something like aether is acknowledged in physics by invoking terms such as "quantum vacuum," "vacuum fluctuations," or "zero-point fluctuations".

V.   **Quantum gravity**. QFTs are generally believed to be a low-energy approximation to a more fundamental theory, cf. [8, Sect.1.3; Chs.11 and 12]. Yet it is not known, nor appreciated what kind of physics is lost as a result of this approximation; e.g. Salam [5, 50] suggested that the physics of quantum gravity is lost. F-medium taking an account of the high-energy properties of the Universe, it might be a useful in modeling quantum gravity, cf. 85].

VI.   **The mean free path of the F-medium suggests a classical explanation for the appearance of some fundamental length in quantum theory**. Already seventy years ago, Heisenberg [22] proposed that quantum mechanics can provide only an idealized, large-scale description of quantum phenomena; there is a kind of fundamental length.

## 5.3   Concluding remarks.

Selection of an appropriate Boltzmann-equation is crucial because it provides a heuristic key to the underlying physics of QFTs. However, as Einstein [51] pointed out: *"The theory we choose decides what we will observe."*

Significant features of Boltzmann-Equation based RTs:

**Feature #1**:  *The QFT free fields* are propagated through the Universe by the F-medium, a relativistic gas [4] consisting of infinitesimal, fermionic and bosonic entities, X-ons, which carry arbitrary four-momentum and interact solely by colliding. We describe their state by

- *The one-particle distribution functions* of the time-space position $x \in \mathbf{R}^{1,3}$ and of the four-momentum $p \in \mathbf{R}^{1,3}$ of gas entities.

- These classical fields of the eight independent, continuous variables $(x, p) \in \mathbf{R}^{1,3} \times \mathbf{R}^{1,3}$ determine the QFT fields, through covariant, local averages over the four-momentum $p$.

- In contrast to the spaces $\mathbf{R}^{1,n}$, $n \geq 4$, of the string theories the physical significance of the eight-dimensional space $\mathbf{R}^{1,3} \times \mathbf{R}^{1,3}$ of RTs is already established, in the kinetic theory of gases.

- The QFT free-field Lagrangians are so modified through Boltzmann-Equations that ultraviolet divergences are eliminated.

**Feature #2:**  *Number of independent variables.* String theories, the Källén–Lehmann spectral representation [8, eq.(10.7.16)], and the *Boltzmann-Equation* approach to a realistic regularization strongly suggest that for modeling the real, physical world with finite QFTs one needs functions of more than four independent variables, i.e. three spatial, one time, and some additional ones. It is still an open question what kind of physical system (framework) we might use to get such additional independent variables. We propose to this end *the F-medium* [4] *and its one-particle distribution functions* of eight independent, continuous variables $\in \mathbf{R}^{1,3} \times \mathbf{R}^{1,3}$.

**Feature #3:**  *Dimensions of the Universe.*  The eight-dimensional space $\mathbf{R}^{1,3} \times \mathbf{R}^{1,3}$ suggests that the Universe has eight dimensions:

1) Four *macroscopic dimensions*: three dimensions of space and one of time; and
2) *Four micro-dimensions: the dimensions* of space $\mathbf{R}^{1,3}$ of the F-medium infinitesimal entities four-momentum.

**Feature #4:**  *Time-scales of microscopic and macroscopic dynamics*. The micro-dynamics of the F-medium is modeled by Boltzmann-Equations and determines the macroscopic dynamics of QFT phenomena. The time-scale for evolution of the macroscopic QFT phenomena apparently bears no relation to the mean collision time, a vastly smaller, basic time-scale for the solutions to Boltzmann-Equation [25]. Within the F-medium based framework for RTs, this disparity presents no such conceptual problems as the hierarchy problems in particle physics [45, 76]. So the F-medium may provide a framework for explaining them. Ross discusses the phenomenological implications of extensions of the standard model capable of solving the hierarchy problem in [82].

**Feature #5:**  *Faster than light* effects suggested by the EPR thought experiments have classical counterparts provided by those phenomena of the F-medium that are spreading faster than the waves; these are customary phenomena of classical gases, cf. [30 ,34, 64].

 **Feature #6:**  *Infinitesimal-range forces*. Whenever we are using some sort of the Boltzmann equation to model the micro-dynamics of a medium, its collision term models, possibly nonlinear and/or infinitesimal-range forces presumed to govern the interactions between the constituent particles of this medium. Thus, if the fundamental partial differential equations of theoretical physics are actually mathematical models of the macroscopic dynamics of a real medium whose micro-dynamics is modeled by a Boltzmann equation, then there might be some novel fundamental forces. We considered how to verify them experimentally in Sect.5.1/v.

**Feature #7:**   *The parameter question*. The Standard Model seems not to be a final theory as there are so many phenomenological parameters. So an underlying theory of QFTs might explain some of them and their numerical values. The proposed realistic regularizations RTs, whose parameters are hypothetical physical constants, might suggest a way to answer this question.

**Feature #8:**   *Unparticles*. F-medium involves directly no particles and consists entirely of X-ons: infinitesimal, fermionic and bosonic entities with no definite mass, which do not interact with electromagnetic radiation. So we could follow Georgi and label the F-medium as "dark unparticle stuff" and X-ons as dark unparticles. About the unparticle stuff described by his theory, Georgi argued in 2007 that it is important to take seriously the possibility that it might actually *exist* in our world [90], i.e. it might be experimentally testable.

### 5.3.1  The basic premises of the paper

**A. Empirical relationships** have been historic stepping stones to theories with physical laws that generalize and extend them, cf. [61; 73, Sect.1]. They and mathematical formulas of physical laws suggested as an inspiring premise the following
**Proposition about the empirical formulas:** *An empirical formula of outstanding fitting quality may be a key stepping-stone towards a pertinent theory*! We can interpret Planck's law, a heuristic key to modern physics, as an extrapolation from the asymptotic behavior for low frequencies toward the one for high frequencies, cf. [73]. Thus we may expect that a realistic regularization of the Standard Model, as an extrapolation of modeling from the finite energies towards infinite ones, will provide a heuristic key to an open Physics beyond the Standard Model, cf. [43, 75, 76, and 81].
**B.   The four premises of  the underlying physics of QFTs:**
   **I.   Lagrangians of QFTs**. *Only the free fields Lagrangians* have to be modified to adequately model the macroscopic, large-scale effects of the micro-properties of the Universe and so avoid ultraviolet divergences. On presuming that the micro-properties of a medium determine its macroscopic properties, the question remains how to model this phenomenon? In 1872 Boltzmann introduced "Kinetic Theory of Gases" with a novel integro-differential equation devised to model the micro-dynamics of fluids; its collision term modeling forces between fluid particles. One can use this equation to derive the partial differential equations that model macroscopic dynamics of fluids. *By using it, Boltzmann undertook to explain the macroscopic properties of dilute gases by analyzing their micro-processes, i.e. the collisions between pairs of molecules* [77].

II.  **Field equations**. Above facts inspired us to propose Boltzmann-Equations for heuristic study into the micro-properties of the Universe, which is based on information about the Universe macroscopic properties modeled by the Standard Model, cf. Sects.1.3 and 5.3.1. By Holman [75]: "there does appear to be broad agreement that some radical renewal, both in concepts and theory, is required in order to move physics beyond the Standard Model." Hundred and fifty years ago Boltzmann's approach was considered as too radical! *Choosing the collision term* for Boltzmann-Equation, one should note what Einstein emphasized about *the significance of theory in observing physical phenomena*. According to Heisenberg [51]: (**a**) "...he insisted that it was the theory, which decides about what can be observed", and (**b**) "Einstein had pointed out to me that it is really dangerous to say that one should only speak about observable quantities. Every reasonable theory will, besides all things, which one can immediately observe, also give the possibility of observing other things more indirectly." The new kind of fundamental forces, the ones between constituent entities of the F-medium might be a case in point.

III. **Matter**. As *physics involves the study of matter*, the question is what kind of matter involves the underlying physics of QFTs. So we presumed a hypothetical F-medium that underlies all physical phenomena modeled by the free field Lagrangians of QFTs. It provides us with a conceptual base for Physics beyond the Standard Model, cf. Sect.5.2.3/5.

IV.  **Planck domain.** That *there is a classical, deterministic theory* for the Planck domain micro-dynamics we are encouraged to presume: **(a)** by 't Hooft's opinion [46]:"It seems quite reasonable first to try a classical, deterministic theory for the Planck domain. One might speculate then that what we call quantum mechanics today, may be nothing else than an ingenious technique to handle this dynamics statistically.", and **(b)** by Feynman's atomistic conjecture that the partial differential equations of motion are actually describing smoothed-out, macroscopic classical motion of some infinitesimal entities [19, Sect.12–7].

### 5.3.2 Premises of the heuristic way to an adequate equation

We now comment on the modifications of a partial differential equation of theoretical physics that are suggested by interpreting it as an approximate model of the macroscopic, large-scale dynamics of a hypothetical medium whose micro-dynamics is modeled by a *Boltzmann-equation*.

 **A. Introduction.** Theoretical physics employs mathematical models of physical phenomena. Of great importance are the models in terms of partial differential equations. Any current mathematical model is most likely a simplification and its equations some kind of approximations to the equations of a more accurate model, which may be unknown. Thus, sooner or later any available model may not be adequate any more. When looking for a way to an adequate mathematical model:

- 't Hooft's guidance is published in "About the real way problems are solved in Modern Physics" [66].
- One could also profit to learn about the attitudes of other Nobel laureates by reading their statements, and Nobel and Feynman lectures [6, 11, 13, 19, 50, 61, 67, 74, 78, and 88].
- There is Ockham's razor and the Law of diminishing returns [80].
- To improve on an inadequate model, mathematical physics could suggest a number of different, better approximations.
- We may use heuristics in constructing a scientific theory!
- According to Dirac in 1930, as noted by Lupu-Sax [83]: "The methods of progress in theoretical physics have undergone a vast change during the present century. The classical tradition has been to consider the world to be an association of observable objects (particles, fluids, fields, etc.) moving about according to definite laws of force, so that one could form a mental picture in space and time of the whole scheme. This led to a physics whose aim was to make assumptions about the mechanism and forces connecting these observable objects, to account for their behavior in the simplest possible way. It has become increasingly evident in recent times, however, that nature works on a different plan. Her fundamental laws do not govern the world as it appears in our mental picture in any very direct way, but instead they control a substratum of which we cannot form a mental picture without introducing irrelevancies." So in order to model macroscopic phenomena we introduced a substratum, the F-medium with infinitesimal entities.

• The diffusion equation has various inspiring physical interpretations, modifications, and applications; cf. Ursell [68], who presents also "A Microscopic View of Diffusion". This we will supplement in the following paragraph by interpreting the diffusion equation as a model of the macroscopic dynamics of a medium, whose micro-dynamics is modeled by a Boltzmann equation.

**B. The diffusion equation.** Take the partial differential diffusion equation

$$\frac{\partial}{\partial t}\varphi(\boldsymbol{r},t) = D\nabla^2\varphi(\boldsymbol{r},t), \qquad (5.1)$$

where D is the diffusion coefficient. We can derive this equation for scalar field $\varphi(\boldsymbol{r},t)$, by using concentration $c(\boldsymbol{r},t)$, current $\boldsymbol{J}(\boldsymbol{r},t)$, Fick's first law of diffusion, and presuming:

$$\varphi(\boldsymbol{r},t) = c(\boldsymbol{r},t)\,, \qquad (5.2)$$

Fick's first law

$$\boldsymbol{J}(\boldsymbol{r},t) = -D\nabla c(\boldsymbol{r},t)\,, \qquad (5.3)$$

and the continuity equation

$$\frac{\partial}{\partial t}c(\boldsymbol{r},t) = -\nabla\cdot\boldsymbol{J}(\boldsymbol{r},t). \qquad (5.4)$$

These equations imply the diffusion equation (5.1), which is Fick's second law of diffusion.

Within the classical kinetic theory of gasses the diffusion equation (5.1) may be derived as an asymptotic, approximate equation for macroscopic variables: *the number density*

$$\varphi(\boldsymbol{r},t) \equiv \int f(\boldsymbol{r},\boldsymbol{p},t)\ \boldsymbol{d^3p} \qquad (5.5)$$

and *the current*

$$\boldsymbol{J}(\boldsymbol{r},t) \equiv \int vf(\boldsymbol{r},\boldsymbol{p},t)\ \boldsymbol{d^3p}\,, \qquad (5.6)$$

where $f(\boldsymbol{r},\boldsymbol{p},t)$ is the probability density function of seven independent, continuous variables: the position $\boldsymbol{r}\in\mathbf{R}^3$ and the momentum $\boldsymbol{p}\in\mathbf{R}^3$ of the gas particles, and time $t$. The changes of this density may be modeled by the linear integro-differential transport equation

$$\frac{D}{Dt} f = \boldsymbol{S} \, f \, , \qquad\qquad (5.7)$$

where:

**a)** The classical substantial derivative

$$\frac{D}{Dt} f \equiv \frac{\partial}{\partial t} f(\boldsymbol{r}, \boldsymbol{p}, t) + v \cdot \nabla f(\boldsymbol{r}, \boldsymbol{p}, t) \, ; \qquad (5.8a)$$

whereas the relativistic substantial derivative

$$\frac{D}{Dt} f \equiv p^0 \frac{1}{c} \frac{\partial}{\partial t} f(\boldsymbol{r}, \boldsymbol{p}, t) + \boldsymbol{p} \cdot \nabla f(\boldsymbol{r}, \boldsymbol{p}, t) \qquad (5.8b)$$

is used for the relativistic probability density function of the time-space position $x \in \mathbf{R^{1,3}}$ and the four-momentum $(p^0, \boldsymbol{p}) \in \mathbf{R^{1,3}}$ of the gas particles.

**b)** The collision term

$$\boldsymbol{S} \, f \equiv \int I(\boldsymbol{r}, \boldsymbol{p}, \boldsymbol{p_i}, t) \, f(\boldsymbol{r}, \boldsymbol{p_i}, t) \, \boldsymbol{d^3 p_i} - a(\boldsymbol{r}, \boldsymbol{p}, t) f(\boldsymbol{r}, \boldsymbol{p}, t) : \quad (5.9)$$

It models the collisions of gas particles, self-interacting and/or interacting with a host medium, cf. [71].The first RHS term gives particles whose momentum is changing from $\boldsymbol{p_i}$ to $\boldsymbol{p,}$ whereas the second RHS term gives particles whose $\boldsymbol{p}$ momentum is lost. The equation (5.7) with $\boldsymbol{S}$ = 0 is an analog of Newton's First Law that describes free streaming of particles; and it implies the continuity relation (5.4) through relations (5.5) and (5.6).

The integro-differential equation (5.7) equals the Euler-Lagrange equations for the following Lagrangian

$$\mathcal{L} \, (f) \equiv \int f(\boldsymbol{r}, -\boldsymbol{p}, t) \{ \frac{D}{Dt} f - \boldsymbol{S} \, f \} \, d^3 \boldsymbol{p} \, , \qquad (5.10)$$

if

$$I(\boldsymbol{r}, \boldsymbol{p}, \boldsymbol{p_i}, t) = I(\boldsymbol{r}, -\boldsymbol{p_i}, -\boldsymbol{p}, t) \text{ and } a(\boldsymbol{r}, \boldsymbol{p}, t) = a(\boldsymbol{r}, -\boldsymbol{p}, t). \quad (5.11)$$

We pointed out in [69] various algorithms for making equations for computing the first-, second-, and third-order approximations to solutions of a Boltzmann equation in the strong scattering asymptote. Since one of these equations is the diffusion equation, these alternative approximations suggest various improvements on the diffusion equation. Already in 1924, Hilbert showed that certain solutions to a Boltzmann equation can be expanded with respect to a small parameter into a power series whose terms satisfy the partial differential equations of the fluid dynamics, cf. [72]. Thus we make the following proposition.

**C. A heuristic proposition about partial differential equations:** *By interpreting a given partial differential equation as an approximation to an appropriate linear integro-differential transport equation, one may be inspired to various, specific improvements on it or replacements of it.*

Generalizing Newton's first law and the concept of contact forces, we formulated a general Boltzmann equation to interpret the fundamental partial differential equations of theoretical physics (Maxwell's equations, the inhomogeneous wave equations for the Lorentz-gauge potentials, Dirac's equation, the Klein-Gordon equation, Proca's equation, etc.) as modeling the macroscopic dynamics of a medium [72, 27, and 29].

The above, distinct derivations of diffusion equation suggest two distinct ways, we could modify and improve it:

**(a)** *A macroscopic way*, by modifying macroscopic equations (5.3) and/or (5.4); like the historic development of the Standard Model, cf. [74].

**(b)** *A micro-way*, by changing the approximation procedure for deriving it and/or modifying the collision term of the underlying transport equation (5.7). *This micro-approach inspired us!*

**D. Modification of fundamental differential equations:** Theoretical physics uses the field concept for analysis of fundamental forces, and models the field dynamics by relativistic partial differential equations, which are specified by Lagrangians. Maxwell supposed that in the case of an electromagnetic field such a Lagrangian models the deformation of the underlying medium, i.e. the luminifereous ether. Nowadays the Lagrangians of fundamental forces are considered solely as their mathematical models, without reference to any medium.

However, no mathematical model has turned out to be perfectly adequate as yet. Sooner or later we will be looking for physically inspired modifications of the fundamental partial differential equations and their Lagrangians, to improve on them. And if so, it might be practical to follow Bjorken and Drell [15], and assume that these equations are models of the macroscopic dynamics of a hypothetical medium, i.e. of some ether.

To this end, we proposed two, the Universe filling hypothetical media whose micro-dynamics is modeled by a *Boltzmann-Equation*:

**(a)** *A gas made of fermionic and bosonic entities*, in Sect.4.2.2.

**(b)** A solid consisting of isotropic bundles of linear flexible strings, which are coupled at each point by forces and force couples [70].

We formulated the *Boltzmann-Equation* for vector wave phenomena of such solids, and pointed out some solids whose wave phenomena are modeled by Maxwell's equations. Thereby we provided a paradigm for interpreting wave phenomena as the macroscopic, large-scale transport phenomena of a solid medium.

## 6 Summary

To provide a framework for a physically based modification of QFTs to make their momentum space integrals convergent; we hypothesized the F-medium — a relativistic gas made of infinitesimal entities [4] — whose macroscopic, large-scale motion is approximately described by the QFT partial differential equations. The kinetic theory of gases motivated us to use Boltzmann-Equations instead of QFT equations, so as to avoid ultraviolet divergences as conjectured by Pauli in 1949 [2, 14] and suggested by Bjorken and Drell in 1965 [17].

The experimentally determined values of regularization parameters of the perturbative S-matrices open a new window into the properties of the Universe. In this way we may gain additional information at accessible energy scales from collider data! *We propose no new kinds of experiments, only the in-depth evaluation of the high precision, large scale collider data*!

## Acknowledgement

We would like to express our thanks to M. Cavedon, G. 't Hooft, F. F. Gorbatsevich, and S. Ribarič for helpful comments.

*Any comment, reference, suggestion, opinion, viewpoint, and whatever is very welcome!*